\documentclass[10pt,twocolumn,twoside]{IEEEtran} 

\usepackage{hyperref}
\newcommand{\newtext}[1]{{#1}}
\newcommand{\oldtext}[1]{}

\usepackage[noadjust]{cite}
\usepackage[pdftex]{graphicx}
\usepackage[cmex10]{amsmath}
\usepackage{amsthm, amssymb, algorithm, algorithmic, array, verbatim}

\title{Strong Secrecy on the Binary Erasure Wiretap Channel Using Large-Girth LDPC Codes}

\author{Arunkumar~Subramanian\textsuperscript{*},~\IEEEmembership{Student~Member,~IEEE,}
				Andrew~Thangaraj,~\IEEEmembership{Member,~IEEE,}
				Matthieu~Bloch,~\IEEEmembership{Member,~IEEE,}
				Steven~W.~McLaughlin,~\IEEEmembership{Fellow,~IEEE,}%
\thanks{A. Subramanian and S. W. McLaughlin are with the Department of Electrical and Computer Engineering, Georgia Institute of Technology, Atlanta, GA 30332, USA. E-mail: arunkumar@gatech.edu, swm@ece.gatech.edu. Phone: +1-404-385-3383. Fax: +1-404-894-7883.}%
\thanks{A. Thangaraj is with the Department of Electrical Engineering, Indian Institute of Technology Madras, Chennai - 600036, India. E-mail: andrew@iitm.ac.in. Phone: +91-44-2257-4424.}%
\thanks{M. Bloch is with Georgia Tech - CNRS UMI 2958 and with the Department of Electrical and Computer Engineering, Georgia Institute of Technology, 57070 Metz, France . E-mail: matthieu.bloch@ece.gatech.edu. Phone: +33-3-87-20-39-26.}%
\thanks{\textsuperscript{*} Corresponding author}%
}

\newtheorem{lemma}{Lemma}

\newtheorem{theorem}[lemma]{Theorem}
\newtheorem{proposition}[lemma]{Proposition}

\newtheorem{corollary}[lemma]{Corollary}
\theoremstyle{definition}

\begin{document}
\maketitle

\begin{abstract}
	For an arbitrary degree distribution pair (DDP), we construct a sequence of low-density parity-check (LDPC) code ensembles with girth growing logarithmically in block-length using Ramanujan graphs. When the DDP has minimum left degree at least three, we show using density evolution analysis that the expected bit-error probability of these ensembles, when passed through a binary erasure channel with erasure probability $\epsilon$, decays as $\mathcal{O}(\exp(-c_1 n^{c_2}))$ with the block-length $n$ for positive constants $c_1$ and $c_2$, as long as $\epsilon$ is less\oldtext{er} than the erasure threshold $\epsilon_\mathrm{th}$ of the DDP. This guarantees that the coset coding scheme using the dual sequence provides strong secrecy over the binary erasure wiretap channel for erasure probabilities greater than $1 - \epsilon_\mathrm{th}$.
\end{abstract}

\ifCLASSOPTIONpeerreview
\begin{center} \bfseries EDICS Category: CRY-OTHE \end{center}
\fi

\IEEEpeerreviewmaketitle

\newlength{\myFigWidth}
\ifCLASSOPTIONtwocolumn
	\setlength{\myFigWidth}{0.8\columnwidth}
\else
	\setlength{\myFigWidth}{0.6\columnwidth}
\fi

\section{Introduction}
The notion of information-theoretic security on a communication system with a passive eavesdropper was first introduced by Shannon in \cite{Shannon1949}. This model consists of three parties --- Alice, Bob and Eve; Alice wants to convey a secret message $S$ to Bob without revealing it to Eve, who can passively intercept the transmission. Shannon's model involves noiseless communication channels and secret communication can be achieved only if Alice and Bob share an encryption key that is not known to Eve. Alice converts the message $S$ into an $n$-symbol cryptogram $X^n$ using the key $K$ and transmits the cryptogram to Bob. The communication scheme is said to attain \emph{perfect secrecy} if $\mathbb{I}(S; X^n) = 0$. Shannon proved that perfect secrecy is guaranteed only if $\mathbb{H}(K) \geq \mathbb{H}(S)$. In other words, Alice and Bob must share a secret key that is at least as long as the confidential message. 

Wyner~\cite{Wyner1975a} introduced an alternate model called the wiretap channel where communication occurs over noisy channels, with Eve receiving a degraded version of the signal received by Bob. Csisz\'ar-K\"orner~\cite{Csiszar1978} considered a generalization of this model where Eve's reception need not be a degraded version of Bob's received signal. In these models, it is possible to achieve secret communication without using a pre-shared encryption key. This is done by exploiting the fact that the wiretapper's channel is ``noisier'' than the legitimate user's channel.

Since perfect secrecy may not be achievable over the wiretap channel for short block-lengths, Wyner introduced the asymptotic notion of \emph{weak secrecy}. If $Z^n$ is the length-$n$ symbol vector received by Eve, weak secrecy is said to be achieved if the rate of information leakage $\frac{1}{n}\mathbb{I}(S; Z^n)$ vanishes as $n \rightarrow \infty$. The authors of \cite{Wyner1975a, Csiszar1978} calculated the secrecy capacity of the respective channels under the weak secrecy condition. A shortcoming of weak secrecy is that the amount of leaked information can be unbounded even if the rate at which it is leaked goes to zero. Maurer and Wolf~\cite{Maurer2000} highlighted this shortcoming and introduced the notion of \emph{strong secrecy}, which requires that the total amount of leaked information $\mathbb{I}(S; Z^n)$ must vanish as $n \rightarrow \infty$. Though the strong secrecy condition is more stringent, it does not reduce the secrecy capacity \cite{Maurer2000, Csiszar1996}. 

In this paper, we consider the binary erasure wiretap channel (BEWC) model, where Bob's channel is noiseless and Eve's channel is a binary erasure channel (BEC). \newtext{Note that this wiretap model is also called a ``binary-erasure-channel wiretap'' (BEC-WT)~\cite{Liu2007} in literature}\footnote{\newtext{Liu, et al.~\cite{Liu2007} considered a generalized version of our BEWC model where there is a binary-input memoryless symmetric-output wiretap, and called it a type-II wiretap model. The original type-II wiretap model introduced by Ozarow-Wyner~\cite{Ozarow1984} has a noiseless main channel with a fixed number of transmitted bits revealed to the eavesdropper. The eavesdropper is also able to choose which bit locations are revealed to her. These two wiretap models are different --- in \cite{Liu2007}, the wiretapper's channel is memoryless, whereas in \cite{Ozarow1984} it is not.}

\newtext{In \cite{Liu2007}, the BEWC model is called a BEC-WT. Contrastingly, Rathi, et al.~\cite{Rathi2009, Rathi2010} use BEC-WT to denote the wiretap model with a BEC main channel and an independent BEC wiretapper's channel.}
}. The BEWC model is important because other wiretap scenarios can be modeled based on the BEWC. For example, the scenarios with a noiseless main channel and a binary symmetric or an additive white Gaussian noise wiretap channel can be modeled \cite{Liu2007} as degraded BEWCs. \newtext{Moreover, the Erasure Decomposition Lemma~\cite[Lemma 4.78]{ModernCodingTheory} lets us model wiretap systems with a noiseless main channel and an arbitrary binary-input memoryless symmetric-output wiretap as degraded BEWCs. }We employ the forward coding approach to achieve secrecy on our wiretap model. Other approaches, like the ones in \cite{Bennett1995, Maurer2000}, use public discussion on authenticated channels in addition to communication on the wiretap channel to achieve secrecy on the overall system.

In ~\cite{Thangaraj2007}, Thangaraj, et al. proposed using the duals of low-density parity-check (LDPC) codes in a ``coset coding scheme'' \cite{Ozarow1984, Wyner1975a} to achieve weak secrecy on the BEWC. They showed that this scheme achieves weak secrecy over \oldtext{BEWC($\epsilon$) }\newtext{BEWC($\xi$)} for \oldtext{$\epsilon > 1 - \epsilon_\mathrm{th}$}\newtext{$\xi > 1 - \epsilon_\mathrm{th}$}, where $\epsilon_\mathrm{th}$ is the BEC threshold of the LDPC \newtext{code} ensembles under message passing (MP) decoding. As an extension to this result, Suresh, et al.~\cite{Suresh2010} showed that strong secrecy on the BEWC can be achieved using the duals of short-cycle-free LDPC codes in the coset coding scheme. They first show that a sufficient condition for strong secrecy is to have the MP block-error probability decay as $\mathcal{O}(1/n^2)$. Using a stopping set based analysis, they prove that short-cycle-free LDPC \newtext{code} ensembles satisfy this condition. Specifically, they show that strong secrecy is achieved on \oldtext{BEWC($\epsilon$) }\newtext{BEWC($\xi$)} for \oldtext{$\epsilon > 1 - \epsilon_\mathrm{ef}$}\newtext{$\xi > 1 - \epsilon_\mathrm{ef}$}, where $\epsilon_\mathrm{ef}$ is the lower bound on error-floor of LDPC codes defined in \cite{Orlitsky2005}. Since $\epsilon_\mathrm{ef}<\epsilon_\mathrm{th}$, there is a gap between the strong and weak secrecy thresholds for finite-girth ensembles.

The work presented in this paper is based on the LDPC code based coset coding scheme of ~\cite{Thangaraj2007, Suresh2010}. We show that the duals of ``large-girth'' LDPC codes achieve strong secrecy on the BEWC with no gap between the strong and weak secrecy thresholds. We do this by first analyzing the asymptotic behaviour of the bit-error probability estimate given by density evolution as the number of iterations increases monotonically. We then construct irregular Tanner graphs for arbitrary degree distribution pairs such that the girth of these graphs increases logarithmically in the number of vertices. This construction\newtext{, which can potentially create disconnected Tanner graphs,} is based on the large-girth regular Ramanujan graphs constructed by Lubotzky, et al.~\cite{Lubotzky1988}. We show that the LDPC codes based on our graphs have a bit-error probability that closely follows the density evolution estimate for increasing iterations. This property, together with the logarithmic increase in the girth of the underlying graphs, guarantees that the duals of our LDPC codes will achieve strong secrecy on the BEWC. 

In recent work \cite{Mahdavifar2010, Hof2010}, polar codes have been suggested as methods for approaching the secrecy capacity of general degraded and symmetric wiretap channels, of which the erasure wiretap channel is a special case. However, since the \oldtext{LDPC} threshold phenomenon \newtext{of LDPC codes} is observed at shorter block-lengths than polarization, there is enough interest in studying the strong secrecy properties of LDPC \newtext{code} ensembles. Also, the mechanism of security using polar and LDPC codes is different. Security of polar codes is proved using the capacity-approaching properties of these codes. In the case of LDPC \newtext{code} ensembles over erasure wiretap channels, we use the duals of codes with a threshold property that need not be close to capacity.

This paper is organized as follows. In Section \ref{sec:coset-coding-strong-secrecy}, we give a brief introduction of the channel model and the coset coding scheme, and relate the strong secrecy condition to the bit-error probability of the duals of the codes used in the coset coding scheme. In Section \ref{sec:bit-error-asymptotics}, \oldtext{we show that the density evolution bit-error estimate exhibits a double exponential decay as the number of iterations increases}\newtext{we give a brief overview of the density evolution analysis for LDPC codes and state the result regarding the double exponential decay of the density evolution bit-error probability estimate as the number of iterations increases}. We then show that this result translates to strong secrecy on the BEWC using the duals of large-girth regular LDPC codes. In Section \ref{sec:large-girth-graphs}, we provide a quick overview of existing constructions for graphs with good girth. We then describe our construction of large-girth graphs and prove that the duals of the resulting LDPC codes achieve strong secrecy. 

\section{Coset Coding Scheme and Strong Secrecy}\label{sec:coset-coding-strong-secrecy}
\begin{figure}%
\centering
\includegraphics[width=\myFigWidth]{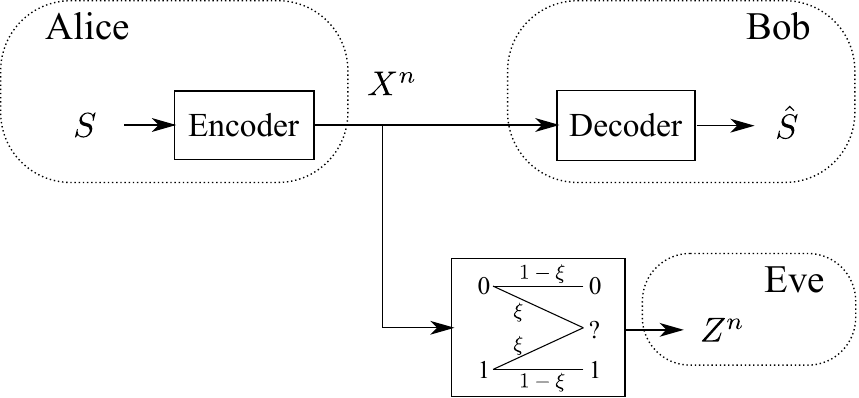}%
\caption{The binary erasure wiretap channel model.}%
\label{fig:bewtc}%
\end{figure}

We consider the binary erasure wiretap channel (BEWC) model introduced in \cite{Thangaraj2007}, which consists of two legitimate parties, Alice and Bob, and an eavesdropper, Eve (Fig. \ref{fig:bewtc}). The channel from Alice to Bob is noiseless and Eve sees the bits sent to Bob through a binary erasure channel (BEC) with erasure probability \oldtext{$\epsilon$ }\newtext{$\xi$}. 

Prior to transmission, Alice and Bob publicly agree on a $(n, n(1-R))$ binary linear code $C$. For each possible value $\mathbf{s}$ of the $nR$ bit secret vector $S$, we associate a coset of $C$ given by $C(\mathbf{s}) = \{ \mathbf{x}^n \in \{0, 1\}^n : \mathbf{x}^n \mathbf{H}^T = \mathbf{s}\}$, where $\mathbf{H}$ is the parity check matrix of $C$. Note that all the vectors in this paper are assumed to be row vectors. To convey the message $S$ to Bob, Alice picks a vector $X^n$ from one of the $2^{n(1-R)}$ vectors of $C(S)$ at random and transmits it over \oldtext{BEWC($\epsilon$) }\newtext{BEWC($\xi$)}. Bob obtains the secret message from the received vector by calculating $X^n \mathbf{H}^T$. 

The amount of information that is leaked to Eve through her observation $Z^n$ can be bounded~\cite{Suresh2010} as 
\newtext{
\begin{align*}
\mathbb{I}(S; Z^n) \leq n\,R\,P_B^\mathrm{MAP}(C^\perp, 1 - \xi)
\end{align*}
}
where \oldtext{$P_B^\mathrm{MAP}(C^\perp, 1 - \epsilon)$}\newtext{$P_B^\mathrm{MAP}(C^\perp, 1 - \xi)$} is the block-error probability under maximum a posteriori (MAP) decoding of the dual code $C^\perp$ transmitted over \oldtext{BEC($1 - \epsilon$) }\newtext{BEC($1 - \xi$)}. A weakened form of this upper bound can be obtained by substituting $P_B^\mathrm{MAP}$ with $P_B^\mathrm{MP}$, the block-error probability using the message passing (MP) decoder with infinite number of iterations. 

The security condition on a sequence of encoding schemes of increasing block-length $n$ and a constant rate $R$ to achieve strong secrecy on a wiretap channel is 
\[
\mathbb{I}(S; Z^n) \rightarrow 0 \quad \text{as } n \rightarrow \infty
\]
\begin{lemma}[adapted from {\cite[Corollary 1]{Suresh2010}}] \label{lem:ber-strong-secrecy-relation}
If $(C^\perp_n)$ is a sequence of binary linear block codes of rate $R$ with increasing block length $n$ such that for some $\alpha > 1$,
\newtext{
\[
P_B^\mathrm{MP}(C^\perp_n, 1 - \xi) = \mathcal{O}\left( \frac{1}{n^\alpha}\right)
\]
}
then strong secrecy is achieved on \oldtext{BEWC($\epsilon$) }\newtext{BEWC($\xi$)} when the dual sequence $(C_n)$ is used under the coset coding scheme.
\end{lemma}

We use the standard order notations $\mathcal{O}$, $o$, $\Theta$, $\omega$ and $\Omega$ as defined in \cite{IntroductionToAlgorithms}. From Lemma~\ref{lem:ber-strong-secrecy-relation}, it is clear that the sequence $(C_n)$ achieves strong secrecy on \oldtext{BEWC($\epsilon$) }\newtext{BEWC($\xi$)} if the dual sequence has 
\newtext{
\[
P_B^\mathrm{MP}(C_n^\perp, 1 - \xi) = \mathcal{O}\left( \frac{1}{n^2}\right) 
\]
}
Using the union bound on the block-error probability,
\newtext{
\[
P^\mathrm{MP}_B(C_n^\perp, 1 - \xi) \leq n P^\mathrm{MP}_b(C_n^\perp, 1 - \xi) 
\]
}
where $P^\mathrm{MP}_b$ is the bit-error probability using the MP decoder. 

\begin{corollary}\label{cor:asymptotic-bit-error-for-strong-secrecy}
If a sequence $(C_n)$ of binary linear codes with increasing block length $n$ and rate $1 - R$ is such that the dual sequence $(C_n^\perp)$ has a bit-error probability such that 
\newtext{
\[
P^\mathrm{MP}_b(C_n^\perp, 1 - \xi) = \mathcal{O}\left( \frac{1}{n^3}\right) 
\]
}
then strong secrecy is achieved on \oldtext{BEWC($\epsilon$) }\newtext{BEWC($\xi$)} when $(C_n)$ is used under the coset coding scheme.
\end{corollary}

\section{Asymptotic Behaviour of the BEC Density Evolution Formula}\label{sec:bit-error-asymptotics}
The asymptotic behaviour of the bit-error probability of LDPC codes \newtext{under MP decoding} can be tracked \oldtext{with} \newtext{using} density evolution~\cite[Sec. 3.9]{ModernCodingTheory}. \newtext{The main result of our paper is based on the asymptotic behaviour of the BEC density evolution expression as the number of iterations goes to infinity. Since density evolution gives only an approximate value of the bit-error probability and our ultimate aim is to study the asymptotic behaviour of the bit-error probability, it is important to understand where and how approximations are made in density evolution.}

\newtext{\subsection{Density Evolution Analysis - a Background}
Let $\mathcal{H}_n$ be an arbitrary ensemble of Tanner graphs with $n$ variable nodes. Suppose a graph $G$ is selected uniformly at random from $\mathcal{H}_n$ and a random codeword from the associated block code is transmitted over BEC($\epsilon$). The receiver, with the knowledge of $G$, tries to decode the transmitted word using the MP algorithm. For a family of ensembles $(\mathcal{H}_n)$ with increasing $n$, let 
	\begin{itemize}
		\item $x(t, n) =$ the probability that a randomly selected edge in the Tanner graph transmits an erasure message from its variable node to its check node at the $t^\mathrm{th}$ iteration
		\item $y(t, n) =$ the probability that a randomly selected codeword bit is unknown after $t$ iterations
	\end{itemize}
}
\newtext{\subsubsection{Computation Graphs~\cite[Sec. 3.7.1]{ModernCodingTheory}}
To evaluate $x(t, n)$ and $y(t, n)$ explicitly, the computation graphs associated with the Tanner graph ensemble may be considered. Suppose a graph $G$ is selected from $\mathcal{H}_n$ uniformly at random and a random edge $e$ is picked from $G$. Let $v$ be the variable node connected to $e$. The level-$t$ \emph{edge-rooted computation graph} $\vec{\mathcal{C}}_t$ of $\mathcal{H}_n$ is defined as the subgraph obtained by traversing from $v$ up to iteration depth $t$ in all directions except along $e$. $\vec{\mathcal{C}}_t$ is a random graph whose distribution depends only on $t$ and $\mathcal{H}_n$. Also, $x(t, n)$ can be uniquely determined given the possible realizations of $\vec{\mathcal{C}}_t$ and their probabilities (regardless of what $\mathcal{H}_n$ is).}

\newtext{To evaluate $y(t, n)$, the level-$t$ \emph{node-rooted computation graph} $\mathring{\mathcal{C}}_t$, defined subsequently, may be considered. As before, a graph $G$ is selected from $\mathcal{H}_n$ uniformly at random. Then, a random variable node $v$ is picked from $G$. $\mathring{\mathcal{C}}_t$ is defined as the subgraph obtained by traversing from $v$ up to iteration depth $t$ in \emph{all} directions. Like $\vec{\mathcal{C}}_t$, $\mathring{\mathcal{C}}_t$ is also dependent only on $t$ and $\mathcal{H}_n$. $y(t, n)$ can be uniquely determined given the possible realizations of $\mathring{\mathcal{C}}_t$ and their probabilities (again, regardless of $\mathcal{H}_n$).}

\newtext{\subsubsection{Tree Ensembles~\cite[Sec. 3.7.2]{ModernCodingTheory}}
While studying the error-correcting performance of LDPC codes, the codes corresponding to the socket permutation ensemble of Tanner graphs, denoted by $\mathcal{G}(n, \lambda, \rho)$, are usually considered. The graphs in this ensemble contain $n$ variable nodes, whose degrees are determined by the degree distribution polynomial $\lambda(x) = \sum_i \lambda_i x^{i-1}$, where $\lambda_i$ is the fraction of edges that are connected to degree-$i$ variable nodes. The check-node degree distribution is determined by the polynomial $\rho(x) = \sum_j \rho_j x^{j-1}$, where $\rho_j$ is the fraction of edges connected to degree-$j$ check nodes. 
}

\newtext{In the classical setting, $\mathcal{H}_n = \mathcal{G}(n, \lambda, \rho)$ is considered, and $x(t, n)$ and $y(t, n)$ are analyzed while keeping $t$ fixed and letting $n$ grow monotonically. The possible computation graphs of $\mathcal{G}(n, \lambda, \rho)$ are not cycle-free and hence enumerating them is cumbersome. Doing an exact analysis of the bit-error probability $y(t, n)$ for this ensemble is therefore difficult. Density evolution resorts to an approximate analysis by considering \emph{tree ensembles}, which are asymptotic approximations of computation graphs.}
	
\begin{figure}
\centering
\includegraphics[width=\myFigWidth]{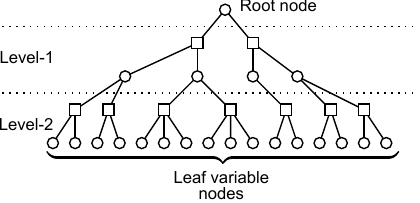}%
\caption{An example of a level-2 decoding neighbourhood tree.}%
\label{fig:decoding-tree-example}%
\end{figure}

	\newtext{The \emph{node-rooted tree ensemble} $\mathring{\mathcal{T}}_t$ is an approximation of the computation graph $\mathring{\mathcal{C}}_t$ of $\mathcal{G}(n, \lambda,\rho)$. $\mathring{\mathcal{T}}_t$ is a random graph which takes all possible level-$t$ decoding neighbourhood trees permitted by $(\lambda,\rho)$ and it is generated by the following rules.
	\begin{itemize}
		\item The degrees of all nodes are chosen independently.
		\item The root variable node has degree $i$ with probability $L_i$, where $L_i$ is the fraction of degree-$i$ variable nodes in $\mathcal{G}(n, \lambda, \rho)$.
		\item All the leaf variable nodes have degree one.
		\item All other variable nodes have degree $i$ with probability $\lambda_i$.
		\item Check nodes have degree $j$ with probability $\rho_j$. 
	\end{itemize}
	The \emph{edge-rooted tree ensemble} $\vec{\mathcal{T}}_t$, which is an asymptotic approximation of $\vec{\mathcal{C}}_t$, is defined in a similar manner, except for the fact that the root variable node has degree $i$ with probability $\lambda_{i+1}$.
	}

	\oldtext{Let $\vec{\mathcal{T}}_t$ be the random variable corresponding to the level $t$ tree ensemble{~\cite[Sec. 3.7.2]{ModernCodingTheory}} corresponding to the DDP $(\lambda, \rho)$.} \newtext{Suppose }\oldtext{We transmit }the \oldtext{``tree code'' }\newtext{block code} corresponding to \newtext{the Tanner graph} $\vec{\mathcal T}_t$ \newtext{is transmitted }over BEC($\epsilon$). The probability that the root node is unknown after $t$ iterations of the MP decoder, denoted by $x_t$, is given by the recursive equation
\[
x_t = \epsilon \lambda( 1 - \rho(1 - x_{t-1})) =: f(\epsilon, x_{t-1})
\]
with $x_0 = \epsilon$. The threshold $\epsilon_\mathrm{th}$ is defined as the supremum of all values of $\epsilon$ for which the sequence $(x_t)$ converges to zero.

For the ensemble $\mathring{\mathcal T}_t$ a similar probability $y_t$ may be defined. We will have
\[
y_t = \epsilon L(1 - \rho(1 - x_{t-1}))
\]
where $L(x) = \sum_i L_i x^i$ is the degree distribution polynomial from the node perspective. 

\newtext{\subsection{Density Evolution - Asymptotic Behaviour}}
It is a well-known result that $x_t$ and $y_t$ exhibit a double-exponential decay as $t$ goes to infinity \newtext{for $\epsilon < \epsilon_\mathrm{th}$}. A proof of this result for regular codes is provided in \cite[Sec. V-A]{Lentmaier2005}. For the sake of completeness, we state the more general result for irregular codes and provide an alternative proof (see Appendix~\ref{proof:de-asymptotics}) involving mathematical induction.

\begin{lemma}\label{lem:de-asymptotics}
	For a distribution pair $(\lambda, \rho)$ with minimum \oldtext{left}\newtext{variable} node degree $l_\mathrm{min} \geq 3$ and $\epsilon < \epsilon_\mathrm{th}$, we have
\begin{align}
x_t, y_t &= \mathcal{O} \left( \exp( - \beta (l_\mathrm{min}-1)^t) \right) \label{eqn:de-asymptotics}
\end{align}
as $t \rightarrow \infty$, where $\beta > 0$ is a constant.
\end{lemma}

\newtext{It is important to note that a similar \emph{double exponential} decay result is not true for DDPs that have degree-2 variable nodes. Working out the expressions for this case, we can see that $x_t$ (and $y_t$) exhibits only an \emph{exponential} decay in the number of iterations. Note that $x_t$ is the expectation of the root-node bit-error probability taken over the possible outcomes of $\vec{\mathcal{T}}$. The dominating term in this expectation is the contribution of the worst-case trees, namely, the trees that contain \emph{only} variable nodes of the least degree. DDPs with degree-2 variable nodes form a special case where the contribution by the worst-case trees decays only exponentially.}

\newtext{\subsection{Asymptotic Decay of Bit-Error Probability}
Suppose we are given a DDP $(\lambda, \rho)$ with $l_\mathrm{min} \geq 3$.} \oldtext{Let the integers $c, d$ be such that $3 \leq c < d$.} For $k = 1, 2, 3, \ldots$, let $(n_k)$ be a strictly increasing sequence of positive integers and let $t_k$ be such that 
\newtext{
\[
t_k = \left\lceil \frac{\log \log n_k + \log a - \log \beta}{\log (l_\mathrm{min} - 1)} \right\rceil
\]
}
for any positive integer $a$. \oldtext{Clearly, t}\newtext{T}his means that $y_{t_k} = \mathcal{O} \left(1/n_k^a \right)$. \newtext{In particular,  we have $y_{t} = \mathcal{O} \left(1/n^3 \right)$ for $a = 3$ (we drop the subscript $k$ for convenience). Since $y_t$ is only an approximation of $y(t, n)$, this does not necessarily mean that the actual bit-error probability $y(t, n)$ itself decays as $\mathcal{O}(1/n^3)$.
}

\newtext{Our ultimate aim is to prove an information-theoretic result and this requires rigorous mathematical proofs. There are only a few rigorous results regarding the ``closeness'' of the density evolution approximation. For example, we know the following results
\begin{itemize}
	\item For $\mathcal{G}(n, \lambda, \rho)$
		\begin{align*}
			\lim_{n \rightarrow \infty} x(t, n) &= x_t,  & 
				\lim_{n \rightarrow \infty} y(t, n) &= y_t
		\end{align*}
		as long as $t$ remains constant~\cite[Thm. 3.49]{ModernCodingTheory}.
	\item The ``exchange of limits'' result by Korada-Urbanke~\cite{Korada2008}.
\end{itemize}
}

\newtext{To achieve strong secrecy, we must find some ensemble $\mathcal{H}_n$ for which $y(t, n) = \mathcal{O}(1/n^3)$, where $t$ is growing with $n$ at least as fast as $\log \log n$. In general, this is not true for $\mathcal{G}(n, \lambda, \rho)$. For example, any irregular DDP with $l_\mathrm{min} = 3$ does not satisfy  $y(t, n) = \mathcal{O}(y_t)$ any $\epsilon > 0$ (see \cite[Thm. 16]{Orlitsky2005}).
}

\subsubsection{Strong Secrecy on the BEWC Using Large-Girth Regular LDPC Codes}
\newtext{Let $\mathcal{G}_{g}(n, \lambda, \rho)$ denote the subset of Tanner graphs in $\mathcal{G}(n, \lambda, \rho)$ whose girth is at least $g$. Clearly, the level-$t$ computation graphs of $\mathcal{G}_{4t+2}(n, \lambda, \rho)$ are cycle free. This means that any possible outcome of $\mathring{\mathcal{C}}_t$ is also a possible outcome of $\mathring{\mathcal{T}}_t$. This does not necessarily mean that $\mathring{\mathcal{C}}_t$ and $\mathring{\mathcal{T}}_t$ are identically distributed and therefore, $y(t, n) = y_t$ is not necessarily true for $\mathcal{G}_{4t+2}(n, \lambda, \rho)$.}

\newtext{The regular LDPC code ensemble $\mathcal{G}_{4t+2}(n, x^{c-1}, x^{d-1})$ is a special case for which $\mathring{\mathcal{C}}_t$ and $\mathring{\mathcal{T}}_t$ are equal to a unique tree $T$. Since $y(t, n)$ is calculated from $\mathring{\mathcal{C}}_t$ in the same way as $y_t$ is calculated from $\mathring{\mathcal{T}}_t$, we have $y(t, n) = y_t$. Using a similar reasoning, we can also say that $x(t, n) = x_t$.
}

\newtext{In essence, density evolution analysis is approximate because it makes the following assumptions. 
\begin{enumerate}
	\item The decoding neighbourhood is a tree
	\item The degrees of the nodes in the decoding tree can be chosen independently.
\end{enumerate}
For large-girth Tanner graphs, the first assumption is justified. However, the second assumption is not justified for large-girth irregular Tanner graphs. For large-girth \emph{regular} Tanner graphs, there is a unique decoding neighbourhood with only one choice for variable-node degrees and only one choice for check-node degrees, which means that the second assumption is also justified. Therefore, we are able to assert that the density evolution estimate is exact in the case of large-girth regular LDPC codes, but are unable to do the same for the irregular counterpart.
}

	Assume that there exists a sequence \oldtext{$(C_{n_k}^\perp)$}\newtext{$(C_{n}^\perp)$} of $(c, d)$-regular LDPC codes \newtext{with $c \geq 3$} such that their Tanner graphs have girth at least \oldtext{$4t_k(3) + 2$ }\newtext{$4t+2$ with
	\begin{align*}
		t = \left\lceil \frac{\log \log n + \log 3 - \log \beta}{\log (c - 1)} \right\rceil
	\end{align*}
	}
	\oldtext{We will prove that such a sequence exists in the next section (more generally, we will prove that asymptotically there exist Tanner graphs of girth $4t_k + 2$ for all integers $a \geq 3$) }\newtext{(The existence of such codes will be proved in the next section)}. For these codes, we have
	\newtext{
\begin{align}
P_b^\mathrm{MP} (C_n^\perp, \epsilon, t) &:= y(t, n) = y_t \label{eqn:regular-asymptotic-bit-error1} \\
				& = \mathcal{O} \left( \frac{1}{n^3} \right) \label{eqn:regular-asymptotic-bit-error}
\end{align}
}
for $\epsilon < \epsilon_\mathrm{th}$\oldtext{ (we drop the subscript $k$ for convenience)}. Here, $P_b^\mathrm{MP} (C_n^\perp, \epsilon, t)$ denotes the bit-error probability after $t$ iterations. By the above equation, the dual sequence $(C_n)$ will achieve strong secrecy on \oldtext{BEWC($\epsilon$) }\newtext{BEWC($\xi$)} for \oldtext{$\epsilon > 1 - \epsilon_\mathrm{th}$}\newtext{$\xi > 1 - \epsilon_\mathrm{th}$}. 

\oldtext{It is important to note that the same argument cannot be used to prove that large-girth \emph{irregular} LDPC \newtext{codes} achieve strong secrecy. This is because \eqref{eqn:regular-asymptotic-bit-error1} is not true for large-girth irregular LDPC codes. Assume that we pick an arbitrary variable node in an irregular Tanner graph of girth at least $4t + 2$ and denote its level $t$ decoding neighbourhood by $N_t$. $N_t$ is a tree because of the lower bound on the girth of the graph. $N_t$ is a random variable that depends on which socket pairing we pick for the Tanner graph. Clearly, the set of all possible values taken by $N_t$ with positive probability is equal to that of $\mathring{\mathcal T}_t$. However, the probability distribution of $N_t$ over the possibilities is different from that of $\mathring{\mathcal T}_t$. Equation \eqref{eqn:regular-asymptotic-bit-error1} is not true for irregular codes for this reason. For regular codes, $N_t$ (and $\mathring{\mathcal T}_t$) can only take one possible value and hence \eqref{eqn:regular-asymptotic-bit-error1} is true. In the next section, we construct an ensemble of large-girth irregular Tanner graphs for which the asymptotic fall of the bit-error probability is the same as that of $y_t$. This aspect is particularly important for secrecy, since we are concerned with the dependence the bit-error probability on the block-length $n$.}

\section{Large-Girth Graphs}\label{sec:large-girth-graphs}
\subsection{Existence of Large-Girth Graphs}
For our scheme to achieve strong secrecy, we require a sequence of regular bipartite graphs whose girth increases faster than $\log \log n$, where $n$ is the number of vertices. We define a sequence of \emph{large-girth graphs} as one with a prescribed degree distribution whose girth increases as $\log n$. These large-girth graphs satisfy the girth condition required for strong secrecy. The existence of these graphs is related to the problem of \emph{cages}~\cite{Wong1982} in graph theory. A $\nu$-regular (simple) graph is one where each vertex has exactly $\nu$ neighbours. A $(\nu, g)$ \emph{cage} is a vertex minimal $\nu$-regular graph of girth $g$. Erd\"os and Sachs~\cite{Erdos1963} showed that cages exist for all $\nu \geq 3$ and $g \geq 3$. For a given $\nu \geq 4$, let $(R_g)_{g \in \mathbb{N}}$ be a sequence of $(\nu, g)$ cages in $n(g)$ vertices. From the upper \cite{Sauer1967} and lower \cite{Erdos1963} lower bounds on $n$, we have $g = \Theta(\log n)$. This means that $(R_g)$ is a sequence of large-girth graphs. Since cages are not necessarily bipartite and we require a sequence of large-girth \emph{bipartite} graphs, we make use of Algorithm \ref{algo:regular-to-bipartite-regular} \cite[Sec. 3.1]{ExtremalGraphTheory}. For convenience, we denote the operation performed by this algorithm by $B(\cdot)$. 

\begin{algorithm}
\caption{Construction of a bipartite graph given any graph \cite[Sec. 3.1]{ExtremalGraphTheory}.}
\label{algo:regular-to-bipartite-regular}
\begin{algorithmic}[1]
	\STATE Given a graph $G$ in $n$ vertices, create an identical copy $G'$ with $V(G) \cap V(G') = \emptyset$. Let $f:V(G) \rightarrow V(G')$ be a graph homomorphism. 
	\STATE Create a graph $H$ with vertex set $V(H) = V(G) \cup V(G')$ and edge set $E(H) = \{ \{x, y\} : x \in V(G), y \in V(G'), f(x) \sim y\text{ in }G'\}$. That is, if $a_1, b_1 \in V(G)$, $a_2 = f(a_1)$, $b_2 = f(b_1)$ and $a_1b_1 \in E(G)$ (or equivalently, if $a_2b_2 \in E(G')$), then $a_1b_2, a_2b_1 \in E(H)$.
\end{algorithmic}
\end{algorithm}

\begin{figure}
\centering
\includegraphics[width=\myFigWidth]{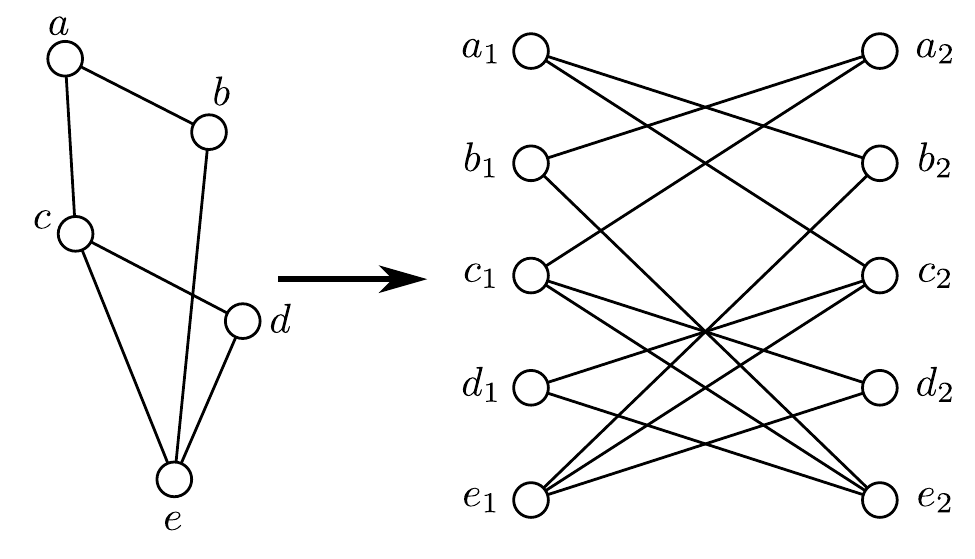}%
\caption{An illustration of Algorithm~\ref{algo:regular-to-bipartite-regular} to create bipartite graphs.}%
\label{fig:bipartite-creation}%
\end{figure}

\begin{lemma}
Given a graph $G$, if $H = B(G)$ then $g(H) \geq g(G)$.
\end{lemma}
\begin{IEEEproof}
For any cycle $C$ in $H$ with the vertices in the order $(u_0, v_0, u_1, v_1, \ldots, u_{r-1}, v_{r-1}, u_0)$ there exists a closed walk 
\[
W = (u_0, f^{-1}(v_0), u_1, f^{-1}(v_1), \ldots, u_{r-1}, f^{-1}(v_{r-1}), u_0)
\] 
in $G$. Note that $r \geq 2$. $W$ can either contain a cycle, or it can be a path tracing itself back after some point. We show that this closed walk has a cycle.

Suppose this closed walk does not have a cycle. Then, without loss of generality, we can assume that it traces itself back at some point $v_i$. Therefore, the sequence $u_i, v_i, u_{(i+1)\bmod r}$ is such that $u_i = u_{(i+1)\bmod r}$. This is a contradiction since all the vertices in the original cycle are distinct and $r \geq 2$. 

Therefore, $W$ contains a cycle $C^*$. We have
\[
\mathrm{length}(C) \geq \mathrm{length}(C^*) \geq g(G)
\]
which proves that $g(H) \geq g(G)$. Note that the second inequality in the above equation follows from the fact that $C^*$ is a cycle in $G$. 
\end{IEEEproof}

By the above lemma, $(B(R_g))_{g \in \mathbb{N}}$ is a sequence of large-girth $\nu$-regular bipartite graphs. Using this sequence, it is possible to construct (see Algorithm \ref{algo:regular-ldpc-large-girth}) large-girth $(c, d)$ regular Tanner graphs for arbitrary $c, d$.

However, it should be noted that there is no standard construction of $(\nu, g)$-cages that works for all $\nu$ and $g$. Moreover, the structures of cages for $\nu > 20$ or for $g > 15$ are not currently known. Though the existence of cages proves the existence of large-girth regular LDPC codes, it does not result in a generalized construction algorithm for these codes. In the later sections, we will construct large-girth LDPC codes from large-girth graphs that are not necessarily cages.

\subsection{Existing Constructions for Tanner Graphs With Good Girth}
A construction of regular large-girth LDPC codes was proposed by Gallager in his monograph \cite[Appendix C]{Gallager1963}. The progressive edge growth (PEG) algorithm \cite{Hu2005} constructs LDPC codes with a prescribed left \newtext{(variable-node)} degree distribution and rate. Though empirical evidence shows that PEG creates codes of good girth, we are unable to prove that they are large-girth codes. A PEG-like algorithm to construct \emph{almost} regular large-girth graphs was proposed in \cite{Chandran2003}. A modification of this algorithm to create almost $(c, d)$-regular large-girth LDPC codes was published as the Almost Regular large-Girth (ARG) algorithm \cite{Krishnan2007}. Though ARG creates large-girth LDPC codes, it cannot be used to create LDPC codes for a pre-defined DDP.

Based on the large-girth regular graph construction in \cite{Lazebnik1995, Lazebnik1995a}, Kim, et al.~\cite{Kim2004} constructed large-girth regular LDPC codes of rate $1/q$, for any prime power $q$. In other work, Margulis~\cite{Margulis1982} constructed $2r$-regular large-girth graphs and based on his idea, Rosenthal and Vontobel~\cite{Rosenthal2001} constructed large-girth (3, 6)-regular LDPC codes using the algebraic structure behind the construction of Ramanujan graphs proposed by \cite{Lubotzky1988}. 

It can be noted that the above constructions produce only large-girth LDPC codes of specific rates and specific (regular) degree distributions. On the other hand, the construction described in the next section produces large-girth graphs of arbitrary rates and degree distributions. 

\subsection{Proposed Construction of Large-Girth Graphs}
The adjacency matrix of a simple graph with $n$ vertices is an $n \times n$ matrix $[a_{i, j}]$ such that $a_{i, j} = 1$ whenever vertices $i$ and $j$ are adjacent, and $a_{i, j} = 0$ otherwise. We consider the eigenvalues of the adjacency matrix. For a $k$-regular graph, any eigenvalue $\mu$ is such that $|\mu| \leq k$. A Ramanujan graph is a $k$-regular graph such that if $\mu$ is an eigenvalue and $|\mu| \neq k$, then $|\mu| \leq 2 \sqrt{k - 1}$. For a detailed discussion of Ramanujan graphs, see the book Davidoff, et al.~\cite{RamanujanGraphs}. 

Lubotzky, et al.~\cite{Lubotzky1988} provided a Cayley graph based construction of certain Ramanujan graphs. For primes $p$ and $q$, they construct a family of graphs $X^{p, q}$ with the following properties. 

\begin{theorem}[{\cite[Thm. 4.2.2]{RamanujanGraphs}}]
Let $p, q$ be distinct, odd primes, with $q > 2\sqrt{p}$. The graphs $X^{p,q}$ are $(p + 1)$-regular graphs that are connected and Ramanujan. Moreover,
\begin{enumerate}
\item If $p$ is a quadratic residue modulo $q$, then $X^{p,q}$ is a non-bipartite graph with $\frac{q(q^2 - 1)}{2}$ vertices, satisfying the girth estimate 
\[
g(X^{p,q}) \geq 2 \log_p q
\]

\item If $p$ is a quadratic non-residue module $q$, then $X^{p,q}$ is a bipartite graph with $q(q^2 - 1)$ vertices, satisfying 
\[
g(X^{p,q}) \geq 4 \log_p q - \log_p 4
\]
\end{enumerate}
\end{theorem}

For our purposes, we will not be \oldtext{needing}\newtext{using} the Ramanujan property of the graphs $X^{p, q}$; we will merely use the above lower bounds on the girth. 

When $p$ is a quadratic residue modulo $q$, we can use the construction in Algorithm \ref{algo:regular-to-bipartite-regular} to generate a bipartite $(p+1)$-regular graph in $q(q^2 - 1)$ vertices with girth at least $2 \log_p q$. Using Algorithm~\ref{algo:regular-to-bipartite-regular}, we now have the following corollary.

\begin{corollary}\label{cor:large-girth-bipartite-prime}
Given a prime $p$, for any $n \in \mathbb{N}$ it is possible to construct a $(p+1)$-regular bipartite graph in $q(q^2 - 1)$ vertices with girth at least $2 \log_p q$ for some prime $q \geq n$ based on the construction of Lubotzky, et al.
\end{corollary}

We would like to construct $k$-regular bipartite graphs of large girth where $k$ is some arbitrary natural number, i.e., it is not necessarily the successor of a prime number. We do this as follows. We first find an integer $s$ such that $sk - 1$ is a prime number, say $p$. The existence of $s$ (and $p$) is guaranteed by the following. 

\begin{theorem}[Dirichlet's Theorem on Arithmetic Progressions {\cite[Chapter 7]{AnalyticNumberTheory}}]
Given two positive integers $a, b$ that are relatively prime, i.e., $\gcd\{a, b\} = 1$, the sequence $(an+b)_{n \in \mathbb{N}}$ contains an infinite number of primes.
\end{theorem}

\begin{corollary}
Given any positive integer $k$, it is always possible to find $s \in \mathbb{N}$ such that $sk - 1$ is a prime. Moreover, there are infinite such $s$. 
\end{corollary}
\begin{IEEEproof}
It can easily be seen that $\gcd\{k, k-1\} = 1$. Therefore, there are infinite prime numbers of the form $r k + (k-1)$, where $r \in \mathbb{N}$. Therefore, there are infinite prime numbers of the form $sk - 1$, where $s \in \mathbb{N}$. 
\end{IEEEproof}

Now, we know that for any arbitrary natural number $k$, we can create a family of $sk$-regular graphs of large girth for some natural number $s$. Using this family, we can create a family of large-girth $k$-regular graphs by using Algorithm~\ref{algo:node-splitting}. 

\begin{figure}%
\centering
\includegraphics[width=\myFigWidth]{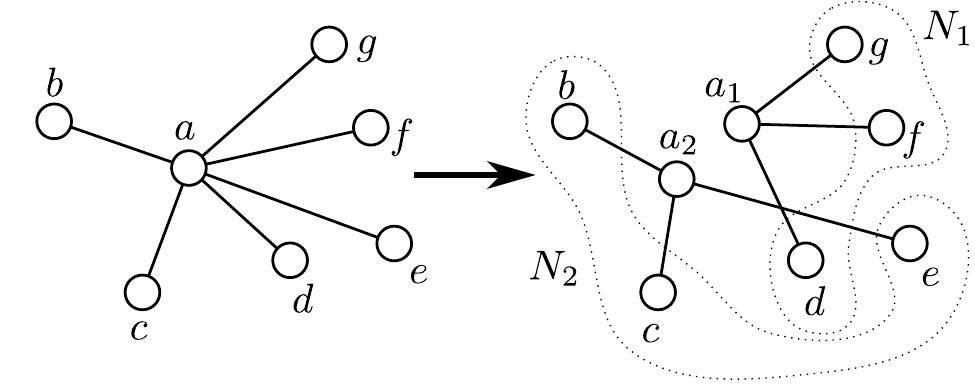}%
\caption{An illustration of Algorithm~\ref{algo:node-splitting} to split a vertex.}%
\label{fig:node-splitting}%
\end{figure}

\begin{algorithm}
\caption{Splitting a vertex into vertices of smaller degrees.}
\label{algo:node-splitting}
\begin{algorithmic}[1]
\STATE \label{item:node-splitting-step1}Given a vertex $v$ in a graph $G$, we partition the set of all its neighbours $N(v)$ into $N_1, N_2, \ldots, N_k$. 
\STATE We create a new graph $H$ by deleting $v$ from $G$ and adding new vertices $v_1, v_2, \ldots, v_k$ and connecting $v_i$ to the vertices in $N_i$ for all $i$. 
\end{algorithmic}
\end{algorithm}
In this paper, we will only consider the creation of equal sized partitions $N_i$ in Step~\ref{item:node-splitting-step1} of Algorithm~\ref{algo:node-splitting}. Under this restriction, the partitioning of $N(v)$ can be done in two different ways.

\begin{itemize}
	\item \textbf{Deterministic version.} We assume that the edges in the graph $G$ are in some simple ordering $(e_1, e_2, \ldots, e_M)$. If $N(v) = \{ e_{i_1}, e_{i_2}, \ldots, e_{i_{jk}}\}$ with $i_1 \leq i_2 \leq \cdots \leq i_j$, let
		\begin{align*}		
			 N_1 &= \{e_{i_1}, e_{i_2}, \ldots, e_{i_{k}}\} \\
			 N_2 &= \{e_{i_{k+1}}, e_{i_{k+2}}, \ldots, e_{i_{2k}}\}
		\end{align*}
		and so on.
\item \textbf{Random version.} The partitioning of $N(v)$ is done in a random fashion.
\end{itemize}
Though the girth properties of the graphs obtained by the deterministic and the random versions of Algorithm~\ref{algo:node-splitting} are similar, it is easier to count graphs of a particular configuration when we use the deterministic version than when we use the random version. 

\begin{lemma}\label{lem:node-splitting-girth}
Given a graph $G$, if $H$ is a graph obtained by splitting an arbitrary vertex of $G$ according to either version of Algorithm~\ref{algo:node-splitting}, then $g(H) \geq g(G)$.
\end{lemma}
\begin{IEEEproof}
	\newtext{Suppose $H$ does not have any cycles. In this case, $g(H) = \infty$ and the lemma is true.}

	\newtext{We are now left with the case where $H$ has cycles.}
Consider any cycle $C$ in $H$. Let $v$ be the vertex of $G$ that is being split and let $V_\mathrm{new} = \{v_1, v_2, \ldots, v_k\}$ be the set of new vertices created. By traversing along $C$ and identifying vertices $v_i$ with $v$, we will get a closed walk $W$ in $G$. We show that $W$ contains a cycle.

If $C$ has less than two vertices from the set $V_\mathrm{new}$, then $W$ is a cycle and we are done. Otherwise, $C$ has at least two of these vertices. We can pick $v_i, v_j$ such that while traversing from one to the other along $C$, we don't encounter any other vertices from $V_\mathrm{new}$. Let this path (excluding $v_i, v_j$) be denoted by $P$. Since $v_i$ and $v_j$ are not adjacent and $N(v_i) \cap N(v_j) = \emptyset$, this path has at least two vertices. Therefore, $vPv$ is a cycle $C^*$ in $W$ that is smaller than $C$. Since $G$ contains the cycle $C^*$, we have 
\[
\mathrm{length}(C) \geq \mathrm{length}(C^*) \geq g(G)
\]
which shows that $g(H) \geq g(G)$.
\end{IEEEproof}

\newtext{Note that Algorithm~\ref{algo:node-splitting} can sometimes create a disconnected graph. That is, $H$ may be disconnected even if $G$ is connected. However, we can see that Lemma~\ref{lem:node-splitting-girth} is valid regardless of any disconnections introduced by node splitting. Furthermore, the proof of our main theorem relies only on Lemma~\ref{lem:node-splitting-girth} and is valid even though some of the Tanner graphs in our ensemble may be disconnected. 
}

\begin{algorithm}
\caption{Constructing large-girth $k$-regular bipartite graphs.}
\label{algo:regular-bipartite}
\begin{algorithmic}[1]
\STATE Given a positive integer $k$, find the smallest solution for $s \in \mathbb{N}$ such that $sk - 1$ is a prime. The existence of $s$ is guaranteed by Dirichlet's Theorem on Arithmetic Progressions. Denote $sk - 1$ by $p$. 
\STATE Pick a sequence of primes greater than $2 \sqrt{p}$. For each such prime $q$, generate the graph $X^{p, q}$ described in \cite{Lubotzky1988}. 
\STATE If $p$ is a quadratic residue modulo $q$, then $G = B(X^{p, q})$. Otherwise, $G = X^{p, q}$. In either case, $G$ is an $(sk)$-regular bipartite graph on $q(q^2 - 1)$ vertices and $g(G) \geq 2 \log_p q$.
\STATE Split each vertex of $G$ successively into $s$ vertices of degree $k$ according to either version of Algorithm~\ref{algo:node-splitting}. Call this new graph $H$.
\STATE $H$ is a $k$-regular bipartite graph with $sq(q^2 - 1)$ vertices and $g(H) \geq g(G) \geq 2 \log_p q$.
\end{algorithmic}
\end{algorithm}

\begin{algorithm} 
\caption{Constructing large-girth $(c, d)$-regular bipartite graphs.}
\label{algo:regular-ldpc-large-girth}
\begin{algorithmic}[1]
	\STATE Let $k = \mathrm{LCM}\{c, d\}$. Construct a sequence of $k$-regular bipartite graphs of large girth according to Algorithm \ref{algo:regular-bipartite}.
	\STATE Given a $k$-regular bipartite graph $G$ with $sq(q^2 - 1)$ vertices and $g(G) \geq 2 \log_p q$, let $(V_s, V_c)$ be the bipartition of the vertices. \newtext{We have $|V_s| = |V_c| = \frac{sq(q^2 - 1)}{2}$.}
	\STATE Split each vertex in $V_s$ into $k/c$ new vertices of degree $c$ each according to either version of Algorithm~\ref{algo:node-splitting} \newtext{to get $\frac{k}{c}\frac{sq(q^2 -1)}{2}$ left vertices of degree $c$}. 
	\STATE Split each vertex in $V_c$ into $k/d$ new vertices of degree $d$ each according to either version of Algorithm~\ref{algo:node-splitting} \newtext{to get $\frac{k}{d}\frac{sq(q^2 -1)}{2}$ right vertices of degree $d$}. The resultant graph $H$ is a $(c, d)$-regular bipartite graph with\oldtext{ bipartition $(V^*_s, V^*_c)$. $H$ has $\frac{sqk}{c}(q^2 - 1)/2$ left nodes of degree $c$ and $\frac{sqk}{d}(q^2 - 1)/2$ right nodes of degree $d$ and} $g(H) \geq 2\log_p q$. 
\end{algorithmic}
\end{algorithm}

\begin{algorithm}
\caption{Constructing large-girth $(\lambda, \rho)$ irregular bipartite graphs.}
\label{algo:irregular-ldpc-large-girth}
\begin{algorithmic}[1]
	\STATE Let $k$ be the least common multiple of all the left and right degrees. Let $a$ be the smallest positive integer such that $a \lambda_i, a \rho_j \in \mathbb{N}$ for all $i, j$.
	\STATE Let $s$ be the smallest natural number such that $sak - 1$ is a prime number. Call this prime number $p$. Choose an arbitrarily prime $q > 2 \sqrt{p}$. Construct an $(ak)$-regular bipartite graph $G_0$ according to Algorithm \ref{algo:regular-bipartite}. $G_0$ has  $sq(q^2 - 1)$ vertices and $g(G_0) \geq 2 \log_p q$.
	\STATE \label{item:irregular-ldpc-large-girth-regular-step} Split each vertex of $G_0$ into $a$ vertices of degree $k$ by successively applying Algorithm \ref{algo:node-splitting} (either version) and denote the resulting $k$-regular bipartite graph by $G$. $G$ has \oldtext{$2n_0 = asq(q^2 - 1)$ vertices }\newtext{$n_0$ vertices on the left and $n_0$ vertices on the right, where $n_0 = \frac{asq(q^2 - 1)}{2}$,} and $g(G) \geq 2 \log_p q$.
	\STATE Let $(v_1, v_2, \ldots, v_{n_0})$ be some ordering of the ``left'' vertices in $G$ and let $(c_1, c_2, \ldots, c_{n_0})$ be some ordering of the ``right'' vertices in $G$. Also, let $(e_1, e_2, \ldots, e_{n_0 k})$ be some ordering of the edges in $G$. 
	\STATE Let $\sigma$ and $\pi$ be two randomly chosen permutation functions over the set $\{1, 2, \ldots, n_0\}$.
	\STATE Consider the ordered set $(v'_1, v'_2, \ldots, v'_{n_0})$, where $v'_i = v_{\sigma(i)}$. In this ordered set, 
	\begin{itemize}
		\item split the first $n_0 \lambda_{l_\mathrm{min}}$ vertices into $n_0 k \lambda_{l_\mathrm{min}}/l_\mathrm{min}$ vertices of degree $l_\mathrm{min}$,
		\item split the next $n_0 \lambda_{l_\mathrm{min}+1}$ vertices into $n_0 k \lambda_{l_\mathrm{min}+1}/(l_\mathrm{min}+1)$ vertices of degree $l_\mathrm{min} + 1$,
		\item $\cdots$
		\item split the last $n_0 \lambda_{l_\mathrm{max}}$ vertices into $n_0 k \lambda_{l_\mathrm{max}}/l_\mathrm{max}$ vertices of degree $l_\mathrm{max}$.
	\end{itemize}
	In the above, we split the vertices according to the deterministic version of Algorithm~\ref{algo:node-splitting}. 
	\STATE Do a similar operation for the check nodes using the ordered set $(c'_1, c'_2, \ldots, c'_{n_0})$, where $c'_j = c_{\pi(j)}$, and the distribution $\rho$. The resulting graph $H$ is a $(\lambda, \rho)$ irregular bipartite graph with 
\[
n = \frac{aksq(q^2 - 1)}{2}\int_0^1 \lambda \mathrm{d}x
\]
 vertices and girth at least $2 \log_p q$.
\end{algorithmic}
\end{algorithm}

\subsection{Strong Secrecy Using Irregular LDPC Codes Based on Ramanujan Graphs}
For a given DDP $(\lambda, \rho)$, we can create a sequence of large-girth $(\lambda, \rho)$-irregular LDPC codes $(C_n)$ of increasing block-length $n$ using Algorithm~\ref{algo:irregular-ldpc-large-girth}. We denote the large-girth graphs associated with $C_n$ by $\mathcal{R}_n$. 

\begin{theorem}
\label{thm:main-thm}
For a given DDP $(\lambda, \rho)$ with minimum left degree $l_\mathrm{min} \geq 3$, the sequence of large-girth $(\lambda, \rho)$-irregular LDPC codes $(C_n)$ created using Algorithm~\ref{algo:irregular-ldpc-large-girth} is such that whenever $\epsilon < \epsilon_\mathrm{th}$ we have 
\begin{align}
\mathbb{E}P_b^\mathrm{MP}(C_n, \epsilon) = \mathcal{O}\left( \exp(- c_1 n^{c_2} )\right)
\label{eqn:main-thm-double-exp}
\end{align}
for some positive constants $c_1, c_2$. 
\end{theorem}

\begin{IEEEproof}
	See Appendix~\ref{proof:main-thm}
\end{IEEEproof}

\oldtext{This}\newtext{The} asymptotic decay \newtext{of the bit-error probability achieved by the codes in Thm.~\ref{thm:main-thm}} is faster than the inverse cubic decay required for strong secrecy\oldtext{ and}\newtext{. T}\oldtext{t}his directly implies that the duals of our Ramanujan graph LDPC codes achieve strong secrecy on the BEWC under the coset coding scheme.

\subsection{Discussion}
For a given DDP $(\lambda, \rho)$, we have constructed a sequence $({C}_n)$ of large-girth LDPC codes based on Ramanujan graphs. For minimum left degree at least three, we showed that for $\epsilon < \epsilon_\mathrm{th}$, we have 
\[
\mathbb{E} P_b^\mathrm{MP} ({C}_n, \epsilon ) = \mathcal{O}(\exp(-\beta n^{a \log (l_\mathrm{min} - 1)} ))
\]
By Corollary~\ref{cor:asymptotic-bit-error-for-strong-secrecy}, the dual sequence $({C}_n^\perp)$ achieves strong secrecy on \oldtext{BEWC($\epsilon$) }\newtext{BEWC($\xi$)} for \oldtext{$\epsilon > 1 - \epsilon_\mathrm{th}$ }\newtext{$\xi > 1 - \epsilon_\mathrm{th}$}. 

\subsubsection{Difference Between Regular and Irregular Codes}
For \emph{any} large-girth regular LDPC code sequence $({C}_n)$, we have 
\[
\mathbb{P}_b^\mathrm{MP}({C}_n, \epsilon) = \mathcal{O}\left( \exp( -\beta n^{a \log (l_\mathrm{min} - 1)} )\right)
\]
for $\epsilon < \epsilon_\mathrm{th}$. This means that the dual sequence achieves strong secrecy on \oldtext{BEWC($\epsilon$) }\newtext{BEWC($\xi$)} for \oldtext{$\epsilon > 1 - \epsilon_\mathrm{th}$}\newtext{$\xi > 1 - \epsilon_\mathrm{th}$}.

For irregular codes \emph{constructed by Algorithm~\ref{algo:irregular-ldpc-large-girth}}, we have shown that 
\[
\mathbb{E} \left( \mathbb{P}_b^\mathrm{MP}({C}_n, \epsilon) \right) = \mathcal{O}\left( \exp( -\beta n^{a \log (l_\mathrm{min} - 1)} )\right)
\]
for $\epsilon < \epsilon_\mathrm{th}$. This means that \emph{on an average}, the dual sequence achieves strong secrecy on \oldtext{BEWC($\epsilon$) }\newtext{BEWC($\xi$)} for \oldtext{$\epsilon > 1 - \epsilon_\mathrm{th}$}\newtext{$\xi > 1 - \epsilon_\mathrm{th}$}. Moreover, most codes in the dual sequence are concentrated around this average strong secrecy property. This is because the Markov inequality guarantees that, with probability tending to 1, any LDPC code from the ensemble will have probability of bit error smaller than $\frac{1}{n^k}$ (for any positive integer $k$). 

Our result for regular LDPC codes is stronger than that that of irregular LDPC codes. However, irregular LDPC codes are important because they have $1 - \epsilon_\mathrm{th}$ very close to their rate. Therefore, irregular codes are instrumental in achieving a secrecy rate very close to the BEWC secrecy capacity.

\subsubsection{Gap Between Achievable Region and Secrecy Capacity}
For a secret information rate $R$, we are interested in the minimum value of Eve's erasure probability $\epsilon$ for which we can ensure strong secrecy over the BEWC using our scheme. \newtext{Since our proof works only for $l_\mathrm{min} \geq 3$,} \oldtext{T }\newtext{t}his involves finding an optimal DDP of rate $R$ and $l_\mathrm{min} \geq 3$ for which the BEC threshold $\epsilon_\mathrm{th}$ is \oldtext{maximum }\newtext{as high as possible}. It can be noted that $\epsilon_\mathrm{th} < 1 - R$. Most of the capacity achieving DDP sequences require $l_\mathrm{min} = 2$ (e.g., the tornado sequence and the right regular sequence in \cite{Oswald2002}). Therefore, there is a small gap between the strong secrecy rate achievable by our technique and the secrecy capacity of the BEWC. 

For example, when we performed a search using the LDPCOPT online tool~\cite{Urbanke2010} for $R = 0.5$ and $l_\mathrm{min} \geq 3$, we found that the maximum value of $\epsilon_\mathrm{th} = 0.4619$ is achieved by the DDP
\begin{align*}
\lambda(x) &= 0.9043388 x^2 + 0.03300419 x^{16} + 0.01434268x^{17} \\
					&\qquad + 0.03535427 x^{18} + 0.01296008 x^{99} \\
\rho(x) &= x^{10}
\end{align*}
This means that the duals of the LDPC codes constructed using Algorithm~\ref{algo:irregular-ldpc-large-girth} will achieve a strong secrecy rate of $0.5$ over BEWC($\epsilon$) for all $\epsilon > 0.5381$. Note that for $\epsilon$ close to $0.5381$, the secrecy capacity of the BEWC is close to $0.5381$. Our coding scheme will achieve a secrecy rate of $0.5$ over this channel, which is $7\%$ less than the secrecy capacity. 

\section{Conclusion and Future Directions}
In this work, we have constructed LDPC codes whose girth increases logarithmically in block-length using Ramanujan graphs. In contrast to existing large-girth constructions, our construction works for arbitrary irregular degree distribution pairs. To our knowledge, this is the first such construction. We have shown that the duals of these LDPC codes achieve strong secrecy on the binary erasure wiretap channel (BEWC), when their minimum left degree is at least three. To achieve secrecy capacity on the BEWC, we require LDPC \newtext{code} ensembles with degree-2 variable nodes. Since our current proof does not apply to these codes, we must look for new techniques to analyze them. A multiedge-type construction, \newtext{similar to the approach of \cite{Rathi2009, Rathi2010}, }might be required to achieve secrecy capacity on the BEWC. 

In addition, the Ramanujan graph ensemble is interesting in the general area of LDPC coding even without the secrecy application. In particular, one can show (through the relationship between the girth and the stopping distance) that the minimum distance of these codes grows at least as $n^b$, for some $b$ such that $0<b<1$. Further properties of this construction could be explored in future work.

\appendices
\section{Proof of Lemma~\ref{lem:de-asymptotics}}\label{proof:de-asymptotics}
For any $x \in [0,1]$, we have 
\begin{align*}
(1 - x)^{d-1}  &\geq 1 - (d-1) x, \qquad \forall d \in \mathbb{N}\\
\Rightarrow \rho(1-x) &= \sum_{d = 2}^{r_\mathrm{max}} \rho_d (1 - x)^{d-1} \\
		& \geq \sum_{d = 2}^{r_\mathrm{max}} (1 - (d-1)x)\rho_d \\
		& = 1 - (r_\mathrm{avg} - 1) x \\
\Rightarrow 1 - \rho(1 - x) & \leq (r_\mathrm{avg} - 1) x
\end{align*}
where $r_\mathrm{avg}$ is the average check-node degree and $r_\mathrm{max}$ is the maximum check-node degree. For $0 \leq (r_\mathrm{avg} - 1) x \leq 1$,
\begin{align}
f(\epsilon, x) &= \epsilon \lambda( 1 - \rho(1 - x)) \notag \\
			& \stackrel{\mathrm{a}}{\leq} \epsilon \lambda((r_\mathrm{avg} - 1) x) \notag \\
			& = \epsilon \sum_{i = l_\mathrm{min}}^{l_\mathrm{max}} \lambda_i ((r_\mathrm{avg} - 1) x)^{i-1} \notag \\
			& \stackrel{\mathrm{b}}{\leq} \epsilon \sum_{i = l_\mathrm{min}}^{l_\mathrm{max}} \lambda_i ((r_\mathrm{avg} - 1) x)^{l_\mathrm{min}-1} \notag \\
\Rightarrow f(\epsilon, x) & \leq \epsilon ((r_\mathrm{avg} - 1) x)^{l_\mathrm{min} - 1} =: g(\epsilon, x) \label{eqn:de-upper-bound}
\end{align}
Note that (a) follows from the monotonicity of $\lambda(x)$, and (b) follows from \newtext{the given condition} $0 \leq (r_\mathrm{avg} - 1) x \leq 1$. To make the notation easier, let us denote $A = \epsilon (r_\mathrm{avg}-1)^{l_\mathrm{min}-1}$. Since we are operating in the region $\epsilon < \epsilon_\mathrm{th}$ where $(x_t)$ converges to zero, there exists an $R$ such that $A x_R^{l_\mathrm{min}-2} \leq 1$ and $(r_\mathrm{avg} - 1) x_R \leq 1$. The first inequality will be used later in the proof. 

Let us construct a sequence $z_{R+i+1} = g(\epsilon, z_{R+i})$ with $z_R = x_R$. It is immaterial what $z_i$ takes when $i < R$. We then claim that $x_{R+i} \leq z_{R+i}$ for any non-negative integer $i$. We can prove this by induction. The base case is when $i = 0$ and it is true by our choice of $z_R$. Assuming the claim is true for some integer $i \geq 0$, we have
\begin{align*}
x_{R+i+1} = f(\epsilon, x_{R+i}) \leq g(\epsilon, x_{R+i}) \leq g(\epsilon, z_{R+i}) = z_{R+i+1}
\end{align*}
The first inequality is due to Eqn. \eqref{eqn:de-upper-bound} and the second inequality is due to the monotonicity of $g$ and the induction hypothesis. This proves the claim.

We have,
\begin{align*}
z_{R+1} & = A z_R^{l_\mathrm{min} - 1} \\
z_{R+i} & = A^{1 + (l_\mathrm{min} - 1) + (l_\mathrm{min} - 1)^2 + \cdots + (l_\mathrm{min} - 1)^{i - 1}} z_R^{(l_\mathrm{min} - 1)^i} \\
		& = A^\frac{(l_\mathrm{min} - 1)^i - 1}{l_\mathrm{min}-2} z_R^{(l_\mathrm{min} - 1)^i} \\
		& = A^\frac{-1}{l_\mathrm{min}-2} \left( A^\frac{1}{l_\mathrm{min}-2} x_R\right)^{(l_\mathrm{min} - 1)^i} \\
		& = A^\frac{-1}{l_\mathrm{min}-2} \exp \left( 
																			(l_\mathrm{min} - 1)^i \left(  \frac{\log A}{l_\mathrm{min}-2} + \log x_R \right) 
															\right) \\
		& = A^\frac{-1}{l_\mathrm{min}-2} \exp \left( - \alpha_R	(l_\mathrm{min} - 1)^i \right)
\end{align*}
Due to our choice of $R$, $\alpha_R \triangleq \frac{-1}{l_\mathrm{min}-2} \log A - \log x_R $ is positive. In fact, we can choose $R$ arbitrarily large, making $x_R$ arbitrarily small and $\alpha_R$ arbitrarily large. For $t \geq R$, we have
\begin{align*}
x_t & \leq z_t \\
		& = A^\frac{-1}{l_\mathrm{min}-2} \exp \left( - \alpha_R	(l_\mathrm{min} - 1)^{t-R} \right) \\
		& = A^\frac{-1}{l_\mathrm{min}-2} \exp \left( - \frac{\alpha_R}{(l_\mathrm{min} - 1)^R}	(l_\mathrm{min} - 1)^t \right) \\
		& = A^\frac{-1}{l_\mathrm{min}-2} \exp \left( - \beta	(l_\mathrm{min} - 1)^t \right)
\end{align*}
Note that $\beta \triangleq \frac{\alpha_R}{(l_\mathrm{min} - 1)^R}> 0$. Therefore, we have 
\begin{equation*}
x_t = \mathcal{O} \left( \exp( - \beta (l_\mathrm{min} - 1)^t) \right) \quad \text{as }t \rightarrow \infty
\end{equation*}

To prove the second half, we \oldtext{first }note that \newtext{for $x \in [0, 1]$}
\begin{align*}
L(x) 
	&= \sum L_i x^i \leq \sum L_i x^{i-1}\\
	&= \frac{1}{\int_0^1 \lambda(x) \mathrm{d}x} \sum_{i = l_\mathrm{min}}^{l_\mathrm{max}} \frac{\lambda_i}{i}x^{i-1} \\
	&\leq \frac{1}{l_\mathrm{min} \int_0^1 \lambda(x) \mathrm{d}x} \lambda(x) \\
\Rightarrow y_t &\leq \frac{1}{l_\mathrm{min} \int_0^1 \lambda(x) \mathrm{d}x} x_t \\
\Rightarrow y_t &= \mathcal{O} \left( \exp( - \beta (l_\mathrm{min}-1)^t) \right) \qedhere
\end{align*}

\section{Proof of Thm.~\ref{thm:main-thm}}\label{proof:main-thm}
The only sources of randomness in Algorithm~\ref{algo:irregular-ldpc-large-girth} for a given large-girth graph $G$ (at the end of Step~\ref{item:irregular-ldpc-large-girth-regular-step}) are the permutation functions $\sigma$ and $\pi$. The probability distribution of $\mathcal{R}_n$ given $G$ is easier to analyze than that of $\mathcal{R}_n$ when $G$ is not specified. Clearly, \eqref{eqn:main-thm-double-exp} is true whenever 
\begin{align}
\mathbb{E}\left( P_b^\mathrm{MP}({C}_n, \epsilon) | G \right) = \mathcal{O}\left( \exp(- c_1 n^{c_2} )\right)
\label{eqn:main-thm-double-exp1}
\end{align}
is true \emph{uniformly} for all possible $G$ in Step~\ref{item:irregular-ldpc-large-girth-regular-step} of Algorithm~\ref{algo:irregular-ldpc-large-girth}.

Note that $P_b^\mathrm{MP}({C}_n, \epsilon)$ denotes the probability of bit-error after infinite iterations of the MP algorithm (or equivalently, when a stopping set is encountered). This probability is clearly less than the probability of bit-error after a finite number of iterations. Therefore \eqref{eqn:main-thm-double-exp1} is true whenever 
\begin{align}
\mathbb{E}\left( P_b^\mathrm{MP}({C}_n, \epsilon, t(n)) | G \right) \leq A(n) = \mathcal{O}\left( \exp(- c_1 n^{c_2} )\right)
\label{eqn:main-thm-double-exp2}
\end{align}
is true for some function $t(n)$. The role played by the quantity $A(n)$ is to ensure that we are able to upper bound the left hand side uniformly in $G$. We pick $t(n) = a \log n$, where $a > 0$ is such that $g(\mathcal{R}_n) \geq 4 a \log n + 2$. We know that $a$ exists because of the large-girth property of $\mathcal{R}_n$. Let $a_\mathrm{max}$ be the maximum possible value for $a$. 

\oldtext{We prove the following}
\begin{proposition} \label{prop:large-girth-ber-and-DE-relation}
For any $\delta \in (0, 1)$, there exists a natural number $N$ such that for all $n \geq N$ we have
\begin{align}
\mathbb{E}\left( P_b^\mathrm{MP}({C}_n, \epsilon, t(n)) | G \right) \leq \frac{1}{1 - \delta} y_{t(n)}(\epsilon)
\label{eqn:large-girth-ber-and-DE-relation}
\end{align}
where $y_{t(n)}(\epsilon)$ is the quantity defined in Lemma~\ref{lem:de-asymptotics}. 
\end{proposition}
We know from Lemma~\ref{lem:de-asymptotics} that 
\begin{align*}
y_{t(n)}(\epsilon) &= \mathcal{O}\left( \exp( -\beta (l_\mathrm{min} - 1)^{t(n)} )\right) \\
         &= \mathcal{O}\left( \exp( -\beta (l_\mathrm{min} - 1)^{a \log n} )\right) \\
				 &= \mathcal{O}\left( \exp( -\beta n^{a \log (l_\mathrm{min} - 1)} )\right)
\end{align*}
The above equation, along with Proposition~\ref{prop:large-girth-ber-and-DE-relation}, completes the proof of the theorem.

\begin{IEEEproof}[Proof of Proposition~\ref{prop:large-girth-ber-and-DE-relation}]
	\oldtext{For an arbitrary vertex $v$ of $\mathcal{R}_n$, let $N_t$ denote the decoding neighbourhood of $v$ up to $t$ iterations}\newtext{Consider the computation graph $\mathring{\mathcal{C}}_t$ of $\mathcal{R}_n$} (we write $t$ for $t(n)$ to make the notation less cumbersome). \oldtext{Let $\mathring{\mathcal{T}}_t(\lambda, \rho)$ denote random variable corresponding to the associated tree ensemble. }Clearly, $\mathbb{P}(\mathring{\mathcal{C}}_t = T) > 0$ if and only if $\mathbb{P}(\mathring{\mathcal{T}}_t = T) > 0$.

Let $T$ be any valid level-$t$ tree in the sense that $\mathbb{P}(\mathring{\mathcal{T}}_t = T) > 0$. Let $P_e(T, \epsilon)$ be the probability that the root node of $T$ is in error when the tree code associated with $T$ is transmitted over BEC($\epsilon$) and decoded with $t$ iterations of the MP decoder. Note the following two equations
\begin{align*}
y_t(\epsilon) &= \sum \mathbb{P}(\mathring{\mathcal{T}}_t = T) P_e(T, \epsilon)\\
\mathbb{E} \left( P_b^\mathrm{MP}({C}_n, \epsilon, t) | G \right) &= \sum \mathbb{P}(\mathring{\mathcal{C}}_t = T | G) P_e(T, \epsilon)
\end{align*}
From the above, we can see that the proof is complete once we show that for some natural number $N$, we have 
\begin{align*}
	\mathbb{P}(\mathring{\mathcal{C}}_t = T | G) \leq \frac{1}{1 - \delta} \mathbb{P}(\mathring{\mathcal{T}}_t = T)
\end{align*}
for all $n \geq N$. 

Let $T$ be a valid level-$t$ tree with $i_0$ being the degree of the root node. Let this tree have $p_i$ variable nodes of degree $i$ (including the root node, but excluding the leaf nodes) and $q_j$ check nodes of degree $j$. We have
\begin{align}
\mathbb{P}(\mathring{\mathcal{T}}_t = T) = L_{i_0} \lambda_{i_0}^{p_{i_0} - 1} \prod_{i = 3, i \neq i_0}^{l_\mathrm{max}} \lambda_i^{p_i} \prod_{j = 2}^{r_\mathrm{max}} \rho_j^{q_j} \label{eqn:tree-ensemble-probability}
\end{align}

Now, consider $\mathring{\mathcal{C}}_t$. The probability that the root node $v$ has degree $i_0$ is clearly $L_{i_0}$. The $i_0$ edges incident with $v$ in $\mathcal{R}_n$ will correspond to $i_0$ edges in $G$ incident with $u$, the parent node of $v$. Let $b(1), b(2), \ldots, b(i_0)$ be the $i_0$ neighbours of $u$ in $G$ corresponding to those edges. Let $c(1), c(2), \ldots, c(i_0)$ be the daughter nodes in $\mathcal{R}_n$ corresponding to the same edges. The number of ways of choosing the permutation function $\pi$ such that node $c(1)$ has degree $j$ is equal to the number of ways of putting $b(1)$ into a slot that corresponds to degree $j$, which is $n_0 \rho_j$.  Note that these slots are numbered. Here, $n_0 = 2n/\left( k \int_0^1 \lambda \mathrm{d}x\right)$ is the number of left (right) vertices in $G$, where $k$ is the LCM of all the degrees in $(\lambda, \rho)$.

In general, whenever $T$ is a valid level-$t$ tree, we have 
\begin{align}
	\mathbb{P}(\mathring{\mathcal{C}}_t = T) & = L_{i_0} \tbinom{n_0 \lambda_{i_0} - 1}{p_{i_0} - 1} (p_{i_0} - 1)! \notag \\
			& \qquad \times \frac{ \left( n_0 - 1 - \sum_{i=2}^{l_\mathrm{max}} p_i\right)! }{n_0!} \prod_{\substack{i = l_\mathrm{min} \\ i \neq i_0}}^{l_\mathrm{max}}\tbinom{n_0 \lambda_i}{p_i} p_i!  \notag \\
			& \qquad \times \frac{\left( n_0 - \sum_{j=r_\mathrm{min}}^{r_\mathrm{max}} q_j\right)!}{n_0!} \prod_{j=r_\mathrm{min}}^{r_\mathrm{max}}\tbinom{n_0 \rho_j}{q_j} q_j! \label{eqn:neighbourhood-tree-probability}
\end{align}

We note the following inequality.
\begin{align}
&\frac{ \left( n_0 - 1 - \sum_{i=2}^{l_\mathrm{max}} p_i\right)! \left( n_0 - \sum_{j=r_\mathrm{min}}^{r_\mathrm{max}} q_j\right)!}{n_0! \ n_0!} \notag \\
					& \qquad \qquad < \frac{1}{ (n_0 - \sum p_i)^{(\sum p_i) - 1} \quad (n_0 - \sum q_j)^{\sum q_j}}\label{eqn:inequality1}
\end{align}
We also see that 
\begin{align}
&L_{i_0} \tbinom{n_0 \lambda_{i_0} - 1}{p_{i_0} - 1} (p_{i_0} - 1)! \prod_{\substack{i = l_\mathrm{min} \\ i \neq i_0}}^{l_\mathrm{max}}\tbinom{n_0 \lambda_i}{p_i} p_i!  \prod_{j=r_\mathrm{min}}^{r_\mathrm{max}}\tbinom{n_0 \rho_j}{q_j} q_j! \notag \\
& \qquad \qquad < L_{i_0} (n_0 \lambda_{i_0})^{p_{i_0} - 1} \prod_{\substack{i = l_\mathrm{min} \\ i \neq i_0}}^{l_\mathrm{max}} (n_0 \lambda_i)^{p_i} \prod_{j=r_\mathrm{min}}^{r_\mathrm{max}} (n_0 \rho_j)^{q_j} \notag \\
& \qquad \qquad =  n_0^{(\sum p_i + \sum q_j) - 1} L_{i_0} \lambda_{i_0}^{p_{i_0} - 1} \prod_{\substack{i = l_\mathrm{min} \\ i \neq i_0}}^{l_\mathrm{max}} \lambda_i^{p_i} \prod_{j=r_\mathrm{min}}^{r_\mathrm{max}} \rho_j^{q_j} \notag \\
& \qquad \qquad = n_0^{(\sum p_i + \sum q_j) - 1} \mathbb{P}(\mathring{\mathcal{T}}_t = T) \label{eqn:inequality2}
\end{align}
Substituting  \eqref{eqn:inequality1} and \eqref{eqn:inequality2} in \eqref{eqn:neighbourhood-tree-probability}, we get
\begin{align}
	\mathbb{P}(\mathring{\mathcal{C}}_t = T) < \frac{\mathbb{P}(\mathring{\mathcal{T}}_t = T)}{\left( 1 - \frac{\sum p_i}{n_0}\right)^{(\sum p_i) - 1} \left( 1 - \frac{\sum q_j}{n_0}\right)^{\sum q_j} } \label{eqn:neighbourhood-probability-last-step}
\end{align}

The proof is complete once we show that 
\begin{align}
\left( 1 - \frac{\sum p_i}{n_0}\right)^{(\sum p_i) - 1} \left( 1 - \frac{\sum q_j}{n_0}\right)^{\sum q_j} \rightarrow 1 \quad \text{as } n \rightarrow \infty
\end{align}

First, we note that $\sum p_i$ and $\sum q_j$ grow exponentially in $t$. This means that there exist constants $\alpha_1, \alpha_2, \beta_1, \beta_2 > 0$ such that 
\begin{align}
\alpha_1 n^{\beta_1} < \sum p_i, \sum q_j < \alpha_2 n^{\beta_2} \label{eqn:tree-vertex-count-inequality}
\end{align}
We have
\begin{align*}
1 > &\left( 1 - \frac{\sum p_i}{n_0}\right)^{(\sum p_i) - 1} \left( 1 - \frac{\sum q_j}{n_0}\right)^{\sum q_j} \\
		&\qquad \qquad > \left( 1 - \frac{\alpha_2 n^{\beta_2}}{n_0}\right)^{\alpha_2 n^{\beta_2} - 1} \left( 1 - \frac{\alpha_2 n^{\beta_2}}{n_0}\right)^{\alpha_2 n^{\beta_2}} \\
		&\qquad \qquad = \left( 1 - \frac{\alpha_2 n^{\beta_2}}{n_0}\right)^{2 \alpha_2 n^{\beta_2} - 1}
\end{align*}

The proof is complete once we show that 
\begin{align}
\left( 1 - \frac{\frac{1}{2} \alpha_2 k\int_0^1 \lambda \mathrm{d}x}{n^{1 - \beta_2}}\right)^{n^{\beta_2}} \rightarrow 1 \label{eqn:limit2}
\end{align}
For this, we pick the constant $a \in (0, a_\mathrm{max}]$ small enough so that $\beta_2 < 0.5$. Observe that for any $\theta > 1$ and $\alpha > 0$, we have 
\begin{align}
\lim_{n \rightarrow \infty} \left( 1 - \frac{\alpha}{n^\theta}\right)^n = 1
\label{eqn:limit1}
\end{align}
Substituting $m = n^{\beta_2}$ in the left hand side of \eqref{eqn:limit2}, we have 
\begin{align*}
\left( 1 - \frac{\frac{1}{2} \alpha_2 k\int_0^1 \lambda \mathrm{d}x}{m^{(1 - \beta_2)/\beta_2}}\right)^m
\end{align*}
which goes to $1$ as $m \rightarrow \infty$.
\end{IEEEproof}


\bibliographystyle{IEEEtran}
\bibliography{All_references}

\ifCLASSOPTIONpeerreview
\pagebreak
\listoffigures
\fi

\end{document}